\documentclass[twocolumn,floatfix,superscriptaddress,amsmath,showpacs,showkeys,aps,prb]{revtex4}

\usepackage{t1enc}
\usepackage[final]{graphicx}
\usepackage{graphicx}
\usepackage{epsfig}
\usepackage{bm}

\usepackage{color}
\usepackage[normalem]{ulem}

\usepackage{graphics}

\begin{document}

\title{Universal and non-universal neural dynamics on small world connectomes: a finite size scaling analysis}

\author{Mahdi Zarepour}
\affiliation{Instituto de F\'{\i}sica Enrique Gaviola (IFEG-CONICET)\\ Ciudad Universitaria, 5000 C\'ordoba, Argentina}
\author{Juan I. Perotti}
\affiliation{Facultad de
Matem\'atica, Astronom\'{\i}a, F\'{\i}sica y Computaci\'on, Universidad Nacional
de C\'ordoba, Instituto de F\'{\i}sica Enrique Gaviola (IFEG-CONICET)\\ Ciudad Universitaria, 5000 C\'ordoba, Argentina}
\author{Orlando V. Billoni}
\affiliation{Facultad de
Matem\'atica, Astronom\'{\i}a, F\'{\i}sica y Computaci\'on, Universidad Nacional
de C\'ordoba, Instituto de F\'{\i}sica Enrique Gaviola (IFEG-CONICET)\\ Ciudad Universitaria, 5000 C\'ordoba, Argentina}
\author{Dante R. Chialvo}
\affiliation{Center for Complex Systems and Brain Sciences
(CEMSC3), Universidad Nacional de San Mart\'{\i}n, Campus Miguelete, 25 de Mayo y Francia, (1650), San Mart\'{\i}n,
Buenos Aires, Argentina.\\
Consejo Nacional de Investigaciones Cient\'{\i}fcas y Tecnol\'ogicas (CONICET), Godoy Cruz
2290, (1425), Buenos Aires, Argentina.}
\author{Sergio A. Cannas}
\affiliation{Facultad de
Matem\'atica, Astronom\'{\i}a, F\'{\i}sica y Computaci\'on, Universidad Nacional
de C\'ordoba, Instituto de F\'{\i}sica Enrique Gaviola (IFEG-CONICET)\\ Ciudad Universitaria, 5000 C\'ordoba, Argentina}

\date{\today}

\begin{abstract}
Evidence of critical dynamics has been recently found in both experiments and models of large scale brain dynamics. The understanding of the nature and features of such critical regime is hampered by the relatively small size of the available connectome, which prevent  among other things to determine its associated universality class. To circumvent that, here we study a neural model defined on a class of small-world network that share some topological features with the human connectome.  We found that varying the topological parameters can give rise to a scale-invariant behavior belonging either to mean field  percolation  universality class  or having non universal critical exponents. In addition, we found  certain regions of the topological parameters space where the system presents a discontinuous (i.e., non critical) dynamical phase transition into a percolated state. Overall these results shed light on the interplay of dynamical and topological roots of the complex brain dynamics.
\end{abstract}


\maketitle

\section{Introduction}

The study of brain functional activity has revealed the existence of correlated fluctuations and  scale invariance similar to those observed in critical phenomena. Such evidence prompted the conjecture that the large scale organisation of the brain emerges at criticality\cite{beggs,chialvo,mora}. Despite the relevance of many of these findings, the small size of the available connectomes \cite{Haimovici2013} don't allow to determine whether the expected finite-size scaling behavior at criticality holds, as well as the associated universality class. In this work we study the dynamical properties of a model defined on a class of small world networks that share some topological features with the human connectome, for a wide range of values of the parameters that define the associated topology and for different system sizes.  A finite-size scaling analysis allowed us not only for a robust characterization of criticality, but also to estimate critical exponents and therefore to identify the universality class of the critical dynamic claimed to be relevant for the emergence of the large scale organization of brain activity.

The paper is organized as follows. In section II we describe the model as well as the simulation  and finite size scaling methods used. The results are presented in section III and discuss the relevance of the main findings in section IV.

\section{Model and Methods}
\subsection{The model}
This work uses  an adaptation of the neural model presented in Ref.\cite{Haimovici2013} running over a small-world network with a weighted adjacency matrix $w_{ij}$.  To mimic the weights distribution of the human connectome\cite{Haimovici2013,Hangmann2008}, the non null $w_{ij}$ were chosen randomly from an exponential distribution $p(w)=\lambda\, e^{-\lambda\, w}$ with $\lambda=12.5$.

The node dynamics of the neural model respond to  the Greenberg-Hastings cellular automaton\cite{Greenberg-Hastings}, in which each node
 $i$ of the network has associated a three state variable $x_i=0,1,2$, corresponding to the following dynamical states: quiescent  ($x_i=0$), excited  ($x_i=1$) and refractory  ($x_i=2$). The transition rules are the following: if a node at the discrete time $t$ is in the quiescent state $x_i(t)=0$ it can make a transition to the excited state $x_i(t+1)=1$ with a small probability $r_1$ or if $\sum_j w_{ji}\,\delta(x_j(t),1) > T$, where $T$ is a threshold and $\delta(x,y)$ is a Kroenecker delta function; otherwise, $x_i(t+1)=0$. If it is excited $x_i(t)=1$ then it becomes refractory $x_i(t+1)=2$ always. If it is refractory $x_i(t)=2$ then it becomes quiescent  $x_i(t+1)=0$ with probability $r_2$ and remains refractory $x_i(t+1)=2$  with probability $1-r_2$. Following Ref.\cite{Haimovici2013} we set $r_1=10^{-3}$ and $r_2=0.3$\footnote{We made several checks for for different values of $r_2\sim 10^{-1}$ and no qualitative differences were observed}. We kept these values fixed in all the calculations.
\begin{figure}[h]
\begin{center}
\includegraphics[width=0.425\textwidth]{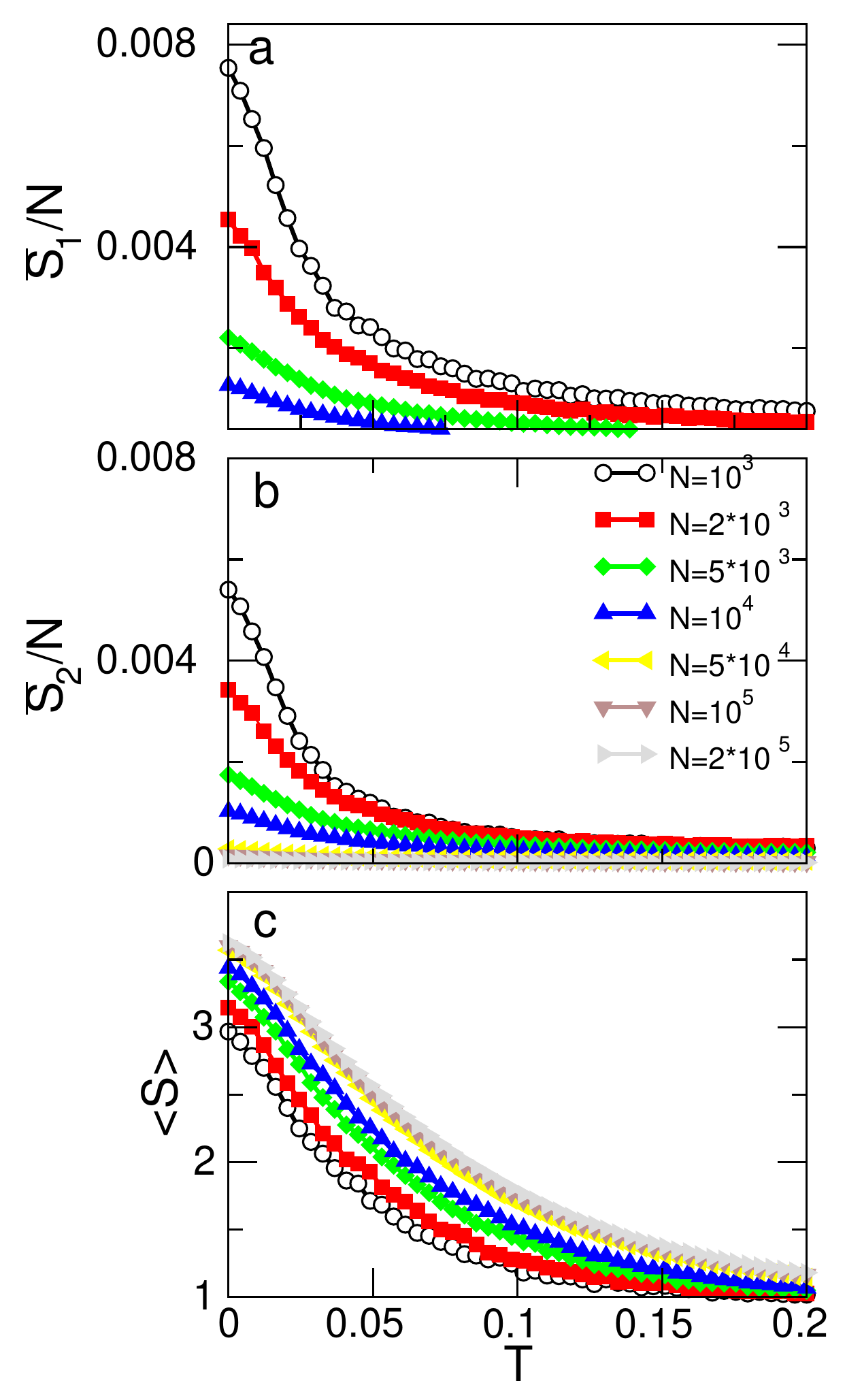}
\caption{\label{fig1} (Color online) Finite size behavior for $\pi=0.6$ and $\langle k \rangle=5$ corresponding to the segregated regime of the model where phase transition is absent for any $T$. ({\bf a}) Order parameter $\overline{S}_1/N$. ({\bf b})  Second largest cluster relative sizes. Notice that they are of the same order of magnitude than the largest clusters shown in Panel A, and that also $\overline{S}_2/N\to 0$ $\forall T$ when $N\to\infty$ ({\bf c}) Average clusters size $\langle s \rangle$ as a function of the threshold $T$ for different sizes $N$. }
\end{center}
\end{figure}
Each node interacts with the others according with the topology of a small-world network, constructed as the usual  Watts-Strogatz model\cite{watts-strogatz}. That is, we start from a ring of $N$ nodes in which each node is connected symmetrically to its $2m$ nearest neighbors. Then, for each node each vertex connected to a clockwise neighbor  is rewired to a random node  with a probability $\pi$ and preserved with probability $1-\pi$, so the average degree $\langle k \rangle=2m$ is preserved\cite{Barrat}.
\subsection{Statistical properties}
The present analysis focuses on the dynamical clusters of coherent activity, namely, groups of nodes simultaneously activated ($x_i=1$) which are linked through non zero weights $w_{ij}$, for different values of the parameters $(\pi,\langle k \rangle)$ and different network sizes $N$. Each simulation started from a random distribution of activated  sites and we let the system to run  $100$ time steps before starting to collect data. We found that time interval to be enough for the system to reach a stationary state for any system size and for any value of the network parameters. Then we recorded data every 5 time steps, to avoid artifactual correlation effects. For each data set we computed several measures  to describe a percolation like critical phenomenon as a function of the control parameter $T$\cite{Haimovici2013}. Specifically, we calculated  the average sizes of  the largest (i.e., giant) cluster $\overline{S}_1$  which can be considered the order parameter and also we compute the average size of the second largest cluster $\overline{S}_2$, together with the average cluster size,

\begin{equation}\label{averages}
    \langle s \rangle = \frac{\sum'_s  s^2 N_s}{\sum'_s  s N_s}
\end{equation}

\noindent where the primed sum runs over all cluster sizes except the giant one and $N_s$ is the number of clusters of size $s$ \cite{Barrat,Margolina} . In some cases we also computed the cumulative complementary distribution function (CCDF) of clusters sizes. Depending of the case,  averages were computed both over data obtained along the same simulation run and over different networks.
\begin{figure}[htb]
\begin{center}
\includegraphics[width=0.425\textwidth]{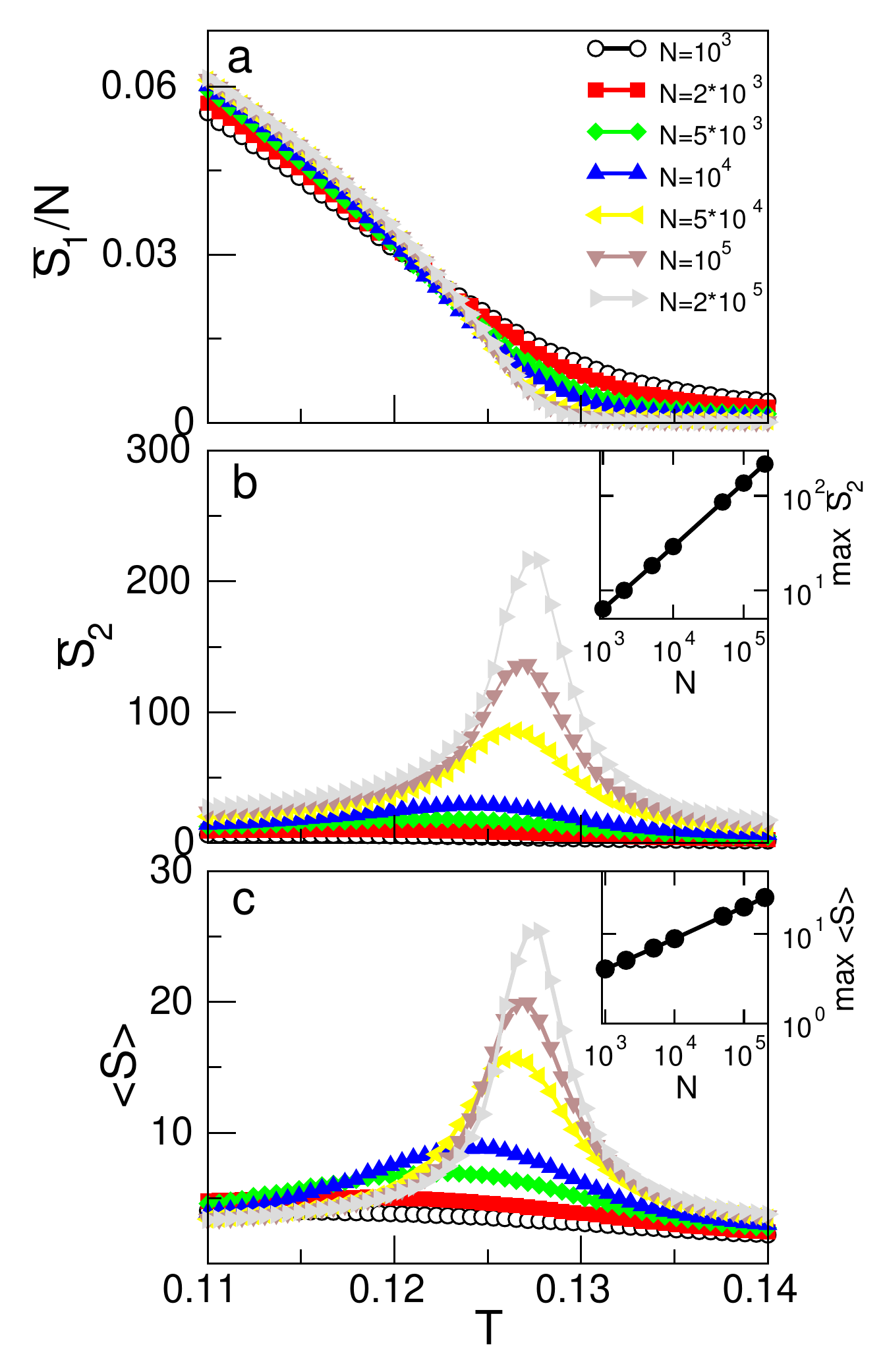}
\caption{\label{fig2} (Color online) Finite size behavior for $\pi=0.6$ and $\langle k \rangle=10$, corresponding to the critical regime (MFC). ({\bf a}) Order parameter. ({\bf b}) Second largest cluster size $\overline{S}_2$. The inset shows the maximum of $\overline{S}_2$ as a function of $N$, where the continuous line is a power law fitting giving a critical exponent $d_f/d=0.35\pm 0.02$. ({\bf c}) Average clusters size $\langle s \rangle$.  The inset shows the maximum of $\langle s \rangle$ as a function of $N$, where the continuous line is a power law fitting giving a critical exponent $\gamma/\nu d=0.67\pm 0.02$.  }
\end{center}
\end{figure}
\begin{figure}[h]
\begin{center}
\includegraphics[width=0.425\textwidth]{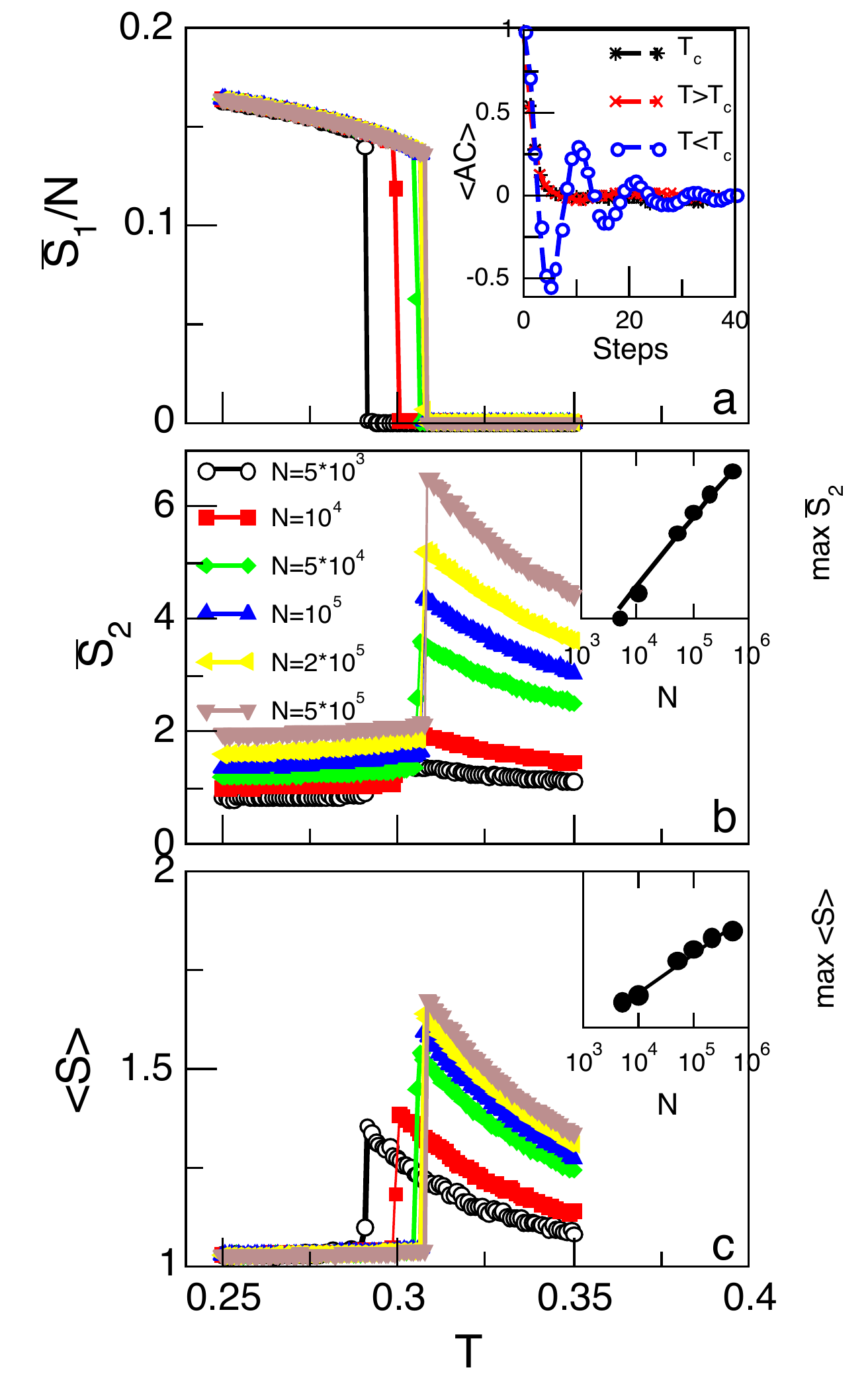}
\caption{\label{fig3} (Color online) Finite size behavior for $\pi=0.6$ and $\langle k \rangle=30$, corresponding to the discontinuous (Disc.) regime. ({\bf a}) Order parameter. The inset shows the autocorrelation function of the average total activity AC close to the critical value $T_c$. ({\bf b}) Second largest cluster size $\overline{S}_2$. The inset shows the maximum of $\overline{S}_2$ as a function of $N$, where the continuous line is a power law fitting giving a critical exponent $d_f/d=0.29\pm 0.02$. ({\bf c}) Average clusters size $\langle s \rangle$.  The inset shows the maximum of $\langle s \rangle$ as a function of $N$, where the continuous line is a power law fitting giving a critical exponent $\gamma/\nu d=0.05\pm 0.02$.  Inset shows the autocorrelation function of the average activity for three values of $T$ very close to $T_c$.}
\end{center}
\end{figure}
If the system presents a percolation like critical point, both $\langle s \rangle$ and   $\overline{S}_2$ are expected to exhibit a (size dependent) maximum for certain pseudo critical value of the control parameter (the threshold $T$ in the present case), that scale with the system size as\cite{Stauffer} $\langle s \rangle \sim N^{\gamma/\nu d}$, $\overline{S}_2 \sim N^{d_f/d}$. $\gamma$ and $\nu$ are the standard susceptibility and correlation length critical exponents, $d$ is the effective dimension of the system and $d_f$ the fractal dimension of the percolating cluster. Also at the critical point it is expected that\cite{Stauffer} $P(s) \sim s^{-\tau} exp({-s/S^*})$,  where $S^* \propto \overline{S}_2$.

\section{Results}

As already mentioned, the  distribution of cluster sizes  of activity can be informative of the different dynamical regimes of the model.
We computed such measures as a function of the threshold $T$ for models of increasing sizes $N$, and different topologies by varying  average degree $\langle k \rangle$ and rewiring $\pi$. The model dynamics can be explored by inspecting the clusters' finite-size scaling behaviour as a function of these three parameters. Figures 1, 2 and 3 correspond to the three typical behaviour found at well defined regions in parameter space which we will describe in detaill now.

At relatively low average degree $\langle k \rangle$, the example of  Fig.\ref{fig1} is typical. We see that  $\overline{S}_1/N$   goes to zero for any value of $T$ as $N \to\infty$.  Also both $\overline{S}_2/N$  and $\langle s \rangle$ display a monotonous decrease (no peak) for any value of the system size $N$. Thus, for those small world topologies only a few small and uncorrelated clusters are present at any time for any value of the threshold $T$, resembling what Tononi\cite{Tononi} describes as \emph{segregated} regime, which per se is incompatible with normal brain functioning.
\begin{figure}[h]
\begin{center}
\includegraphics[width=0.45\textwidth]{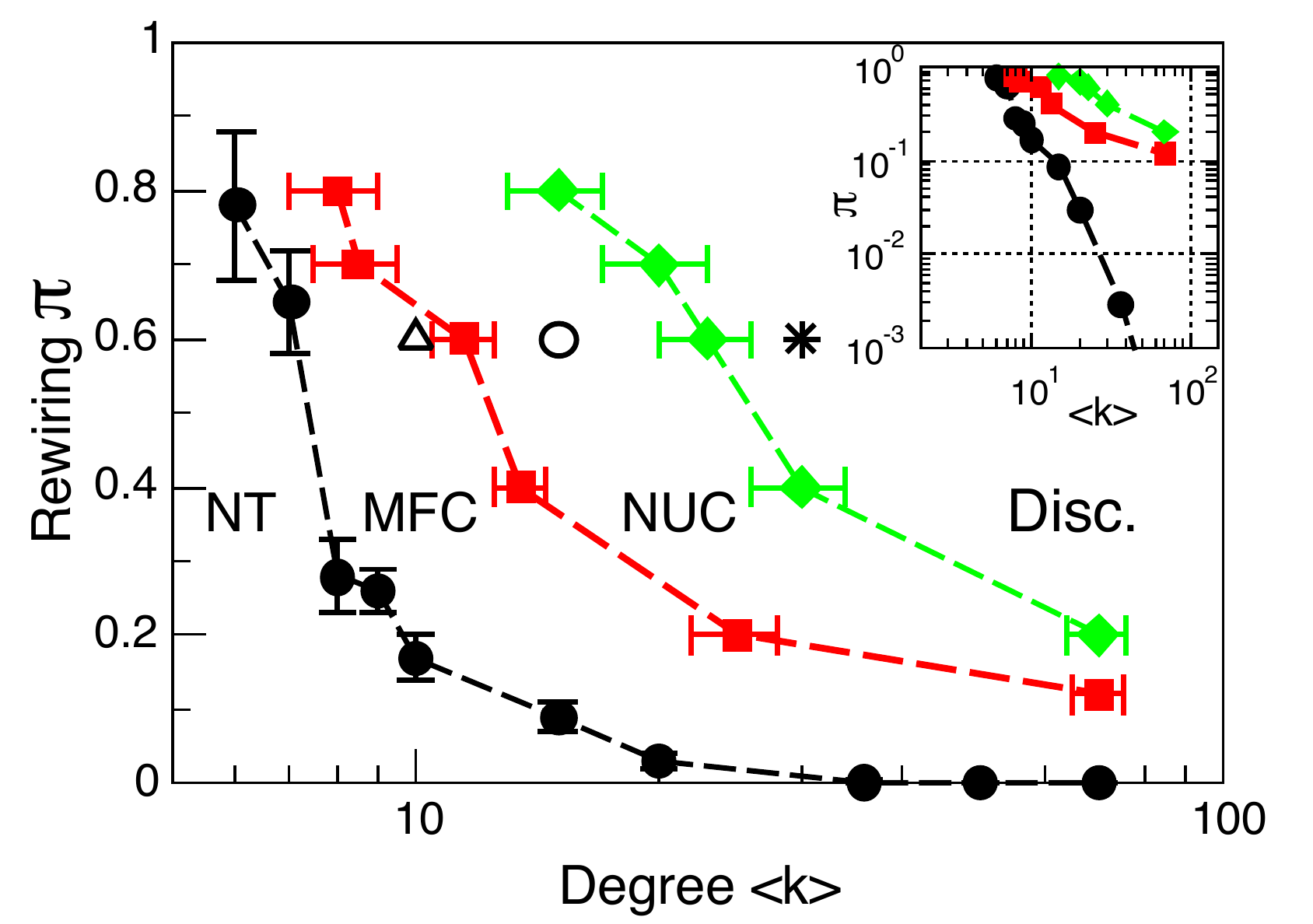}
\caption{ \label{fig4} (Color online)
Parameter space for the network topology values $(\pi,\langle k \rangle)$ in the large $N$ limit. As a function of the node threshold $T$, the system can exhibit a percolation like dynamical phase transition for parameter values above the black line. Below the black line there is no dynamical phase transition (NT). Between the black  line and the red dashed one (MFC)  the behavior became scale-invariant --at a certain critical threshold $T$-- with exponents consistent with the mean field percolation universality class. Further above, in the region denoted as NUC, behavior can be still scale invariant and critical (i.e. the transition is still second order) but without universal exponents.  Finally, above the green dash-dotted line, the transition becomes discontinuous (Disc.) and the dynamic is short-range correlated and oscillatory. The  triangle, circle and star symbols at $\pi=0.6$ correspond to the parameters values used for the statistics presented in Figure 5, Panels A-C respectively. In the inset the same data is plotted in double log axis to best denote the relative sizes of each dynamical regime.
}
\end{center}
\end{figure}
\begin{figure}[h]
\begin{center}
\includegraphics[width=0.45\textwidth]{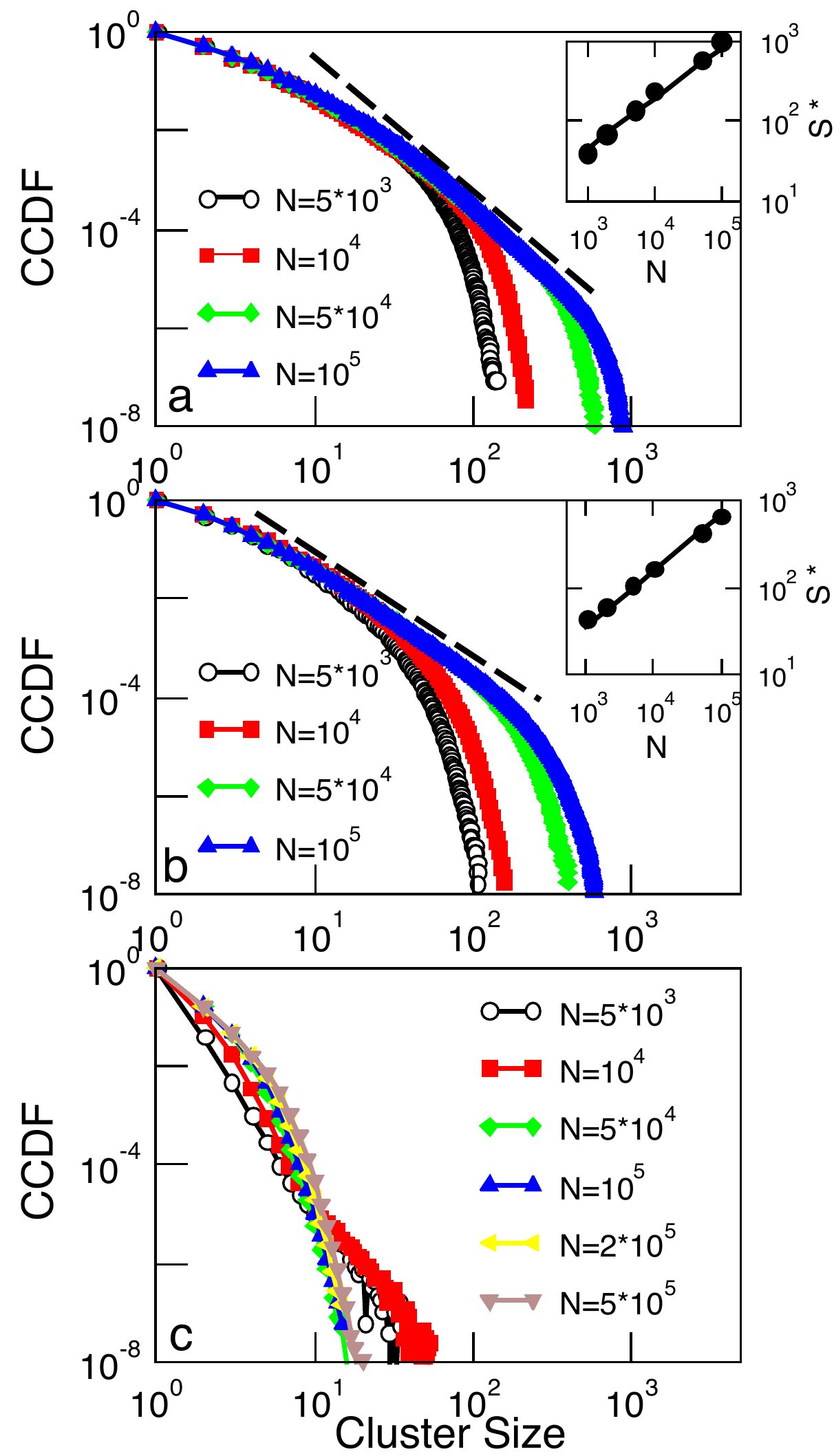}
\caption{ \label{fig5} (Color online)
{\bf (a)} Cumulative cluster size distribution function (CCDF) for $\langle k \rangle=10$, $\pi=0.6$ and different system sizes. The dashed line corresponds to a linear fitting of the central part of the distribution with a power law $\sim s^{-1.58}$. The inset shows the cutoff $S^*$ as function of $N$; the continuous line is a power law fitting giving an exponent $0.62\pm 0.05$.{\bf  (b)} Cumulative cluster size distribution function (CCDF) for $\langle k \rangle=15$, $\pi=0.6$ and different system sizes. The dashed line corresponds to a linear fitting of the central part of the distribution with a power law $\sim s^{-1.54}$. The inset shows the cutoff $S^*$ as function of $N$; the continuous line is a power law fitting giving an exponent $0.60\pm 0.03$.  {\bf (C)} CCDF of the cluster size distribution for $\langle k \rangle=30$, $\pi=0.6$ and different system sizes.
}
\end{center}
\end{figure}
Increasing the value of $\langle k \rangle$ result in a change of the model dynamics. For such case, as shown in Fig.\ref{fig2}, the clusters analysis reveals the characteristic finite-size scaling behavior exhibited by percolation phenomena\cite{Stauffer}. There, we observe the existence of  a critical threshold $T_c$ below which the order parameter converges to a finite value when $N\to\infty$, while it goes to zero above the critical threshold (Fig.\ref{fig2}a). This means the existence of a dynamical phase transition, where the system develops long range order in the form of macroscopic correlated cluster of simultaneously active nodes, which resembles the correlations associated with the brain RSN described in [\cite{Haimovici2013}] . Also both $\overline{S}_2$  and $\langle s \rangle$ exhibit a sharp peak around $T=T_c$ (See Fig.\ref{fig2}b and c) that diverges as a power law when $N\to\infty$, thus exhibiting well defined critical exponents. Hence, for the associated topologies, the dynamical phase transition is indeed a second order or critical one.

A very different transition happens for large enough values of both parameters $(\pi,\langle k \rangle)$. Indeed, in that region both the order parameter and the fluctuations measures $\overline{S}_2$ and $\langle s \rangle$ exhibit an apparently discontinuous behavior at the transition point, as illustrated in Fig.\ref{fig3}. 

By analyzing the finite size scaling of any of the three quantities $\overline{S}_1/N$, $\overline{S}_2$ and $\langle s \rangle$ in the large $N$ limit we estimated  the frontiers in the $(\pi,\langle k \rangle)$ space  between the regions where the system exhibit or o not a dynamical phase transition, as well as the nature of the transition (see Fig.\ref{fig4}).
The frontier between the segregated and the Mean Field Critical (MFC) regimes was obtained by analizing  whether  there is a $T_c>0$ such that $\overline{S}_1/N$  converges to a finite value for $T<T_c$ when $N\to\infty$. The frontier between MFC and the Non Universal Critical (NUC) regime was identified by analyzing whether the difference between  critical exponents  $\gamma/\nu d$ and $d_f/d$ (obtained from the finite size scaling extrapolation) and the corresponding percolation mean field values exceeds or not the associated error bars. The frontier between NUC and the oscillatory regime was estimated by analyzing the presence or not of  a discontinuity in the order parameter at the transition in the large $N$ limit.

At relatively low average degree $\langle k \rangle$ there is no transition (i.e.,  the region labeled NT in Fig.\ref{fig4}). When present, the nature of the dynamical phase transition  can be different depending on the topology of the network, as depicted in Fig.\ref{fig4}.

For relatively small values of $\langle k \rangle$  and/or small enough values of $\pi$, the behavior is critical and the corresponding critical exponents are consistent with the universality class of mean field (or Bethe) classical percolation, namely $\gamma/\nu d\approx 1/3$ and $d_f/d=2/3$ (corresponding to $d=6$, $d_f=4$, $\nu=1/2$ and $\gamma=1$ ). An example is the case already shown in Fig.\ref{fig2}.

To get further insight about the nature of the dynamical phase transition in the last mentioned region, we analyzed the associated behavior of the cluster size distribution (see  Fig.\ref{fig5}). For the critical dynamics, consistently, the cumulative cluster size distribution exhibits the expected behavior $P_c(s) \equiv \sum'_{s'\geq s} P(s') \sim s^{-(\tau-1)} exp({-s/S^*})$, with an exponent $\tau \approx 5/2$ and $S^*\propto  \overline{S}_2 $(thus satisfying the scaling law $\tau=d/d_f+1$)  as shown in Fig.\ref{fig4}, corresponding to the region in the $(\pi,\langle k \rangle)$ space denoted as MCF (Mean Field Critical).

\begin{figure}
\begin{center}
\includegraphics[width=0.45\textwidth]{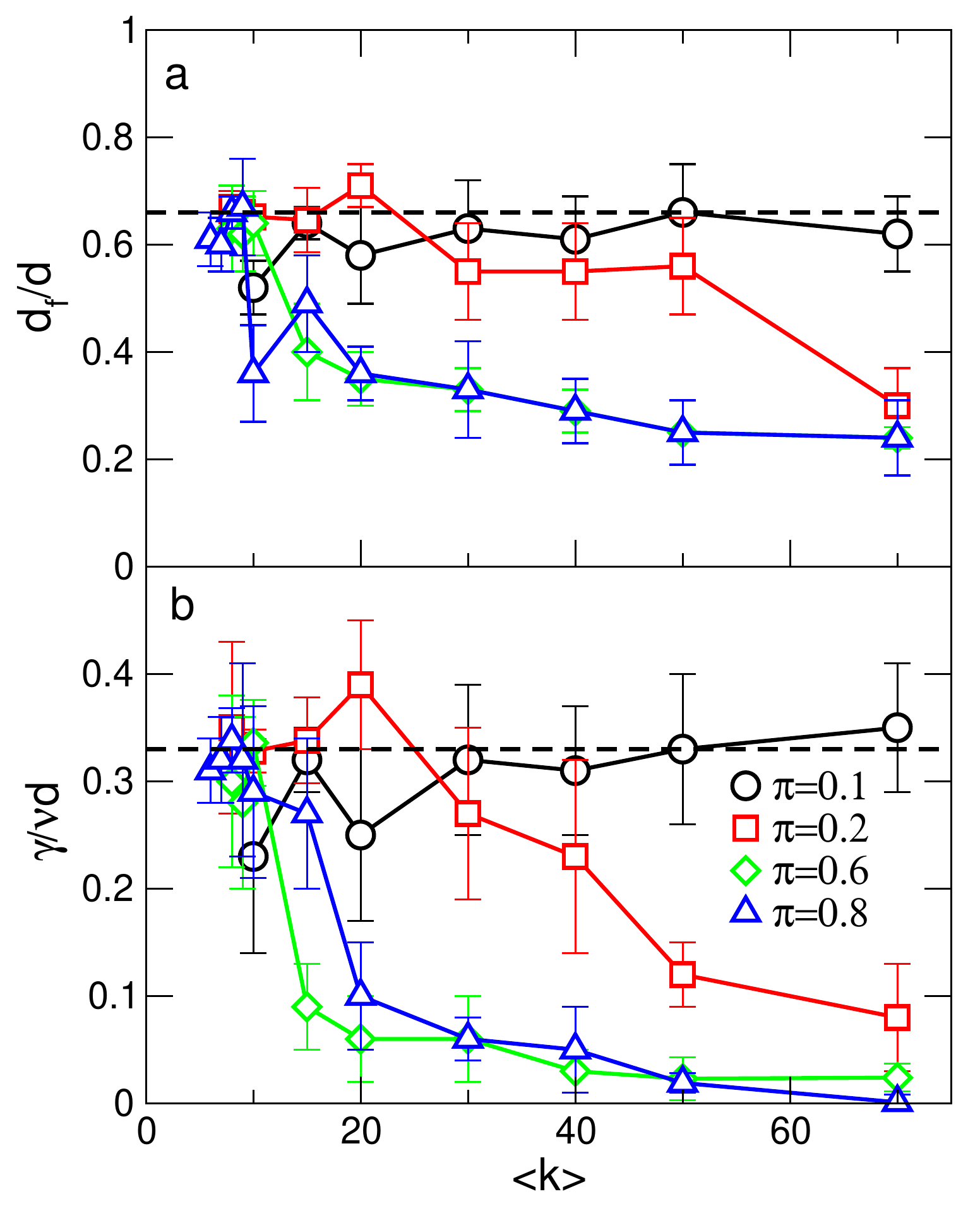}
\caption{\label{fig6} (Color online) Critical exponents obtained by finite size scaling analysis as a function of $\langle k \rangle$ for the range of the rewiring probability $\pi$  values denoted in the legend. ({\bf a}) Critical exponent of the maximum of $\overline{S}_2$. The dashed line corresponds to the mean field value $d_f/d=2/3$. ({\bf b}) Critical exponent of the maximum of $\langle s \rangle$. The dashed line corresponds to the mean field value$\gamma/\nu d\approx 1/3$.}
\end{center}
\end{figure}

For even larger values of $\pi$ and/or $\langle k \rangle$  a wide  region of the parameter space results in critical exponents changing continuously with the topological  parameters, until they seems to saturate (see Fig.\ref{fig6}).
On the other hand, the power law increase of values of $\overline{S}_2$ (see Fig.\ref{fig3}B) at both sides of the discontinuity with an exponent close to one third implies a roughly two dimensional percolating cluster ($d_f\approx 2$), assuming an effective dimension $d=6$.


For comparison, we also calculated the clusters size distribution for a set of values in the NUC region, as shown in Fig.\ref{fig5}B. The  power law distribution is indeed consistent with critical behavior.
On the other hand, the data on Fig.\ref{fig5}C shows that the cluster size distribution for large values of $(\pi,\langle k \rangle)$, for large enough system sizes ($N> 10^4$), exhibits an exponential decay,  which exclude critical behavior. Hence, we can conclude that in such region the system exhibits a first order or discontinuous dynamical percolation transition. In this region the analysis of the autocorrelation function of the mean activity (see Inset Fig. \ref{fig3}a) reveals that the dynamic exhibits a finite time scale corresponding to an oscillation whose period is a function of the topological parameters $\pi$ and $<k>$ as well as $T$.

The results of the scaling analysis are summarized in Fig.\ref{fig6} which show the computed critical exponents both for the maximum of $\overline{S}_2$ in panel A and for the maximum  of $\langle s \rangle$ in panel B. The data shows the two behavior already discussed: the universal (mean field, denoted with filled circles) as well as for the cases in which there is a continuously changing exponent depending on the $\pi$ value.

\section{Discussion}

The main results of the present work  illustrate how the large scale correlated activity depend on the topological details of the underlying structural network.  The results show that the correlated patterns exhibiting scale invariance seen in the brain RSN need certain minimum of  connectivity conditions. In other words, using Tononi's terminology\cite{Tononi},  for  relatively low average degree $\langle k \rangle$ and fraction of network shortcuts $\pi$ the neuronal activity is too segregated.  In the other extreme,  for very large values of $\langle k \rangle$  and $\pi$ the activity is too integrated, exhibiting a first order or discontinuous phase transition. This is most unexpected, because in conventional static percolation on small world networks the transition is second order, belonging to the mean field universality class for any value of the rewiring probability $\pi$ \cite{Moore}. Hence, the present effect is due entirely to dynamical correlations.
 In between these two extremes we found a  regime that seems compatible with the fMRI brain data. There, two different dynamical regimes are found which are characterized by different kinds of dynamical phase transitions as the threshold controlling the neuronal activation is varied. For certain  values the phase transition is universal and for others,  relatively larger, while the scale invariance persist the universality is lost,i.e. the exponents change continuously. Such regions could correspond, with some differences, to the equivalent to  the Griffiths phases\cite{rare1} already described in another context\cite{rare2,moretti}. Please notice some recent results in the context of the brain connectome \cite{odor1,odor2}

From the neuroscience point of view, the relevant finding is that the critical regime in this class of networks spans a wide region in the parameters space, corresponding to intermediate values of networks topologies, those that are too connected or too disconnected are not able to exhibit critical dynamics, regardless of the values of the other two parameters (system size and node's threshold). In other words, from the point of view of neural architecture a minimum of connectivity needs to be predetermined (perhaps via evolution) in order for the dynamics to achieve the dynamical features of criticality (via modulation of excitability or threshold). On the other hand, the discontinuous percolation that appears only for  extremely connected networks,  might constitute pathological conditions observed in real nervous systems. Hence, critical fluctuations emerge as a robust characteristic towards variations in the anatomical network topology which is  consistent with the expectations of criticality described for the spontaneous fluctuations of brain dynamics.

\acknowledgments
This work was partially supported by  CONICET (Argentina) through grant
PIP 11220150100285,   Ministerio de Ciencia y Tecnolog\'{\i}a de la Provincia de C\'ordoba (Argentina) through grant PID2018 and  SeCyT (Universidad Nacional de C\'ordoba, Argentina) and by the NIH (USA) Grant 1U19NS107464-01. MZ is a recipient of a Doctoral Fellowship from CONICET (Argentina). This work used Mendieta Cluster from CCAD-UNC, which is part of SNCAD-MinCyT, Argentina.

\end{document}